\documentclass[aps,prl,twocolumn,grouptaddress,amsmath,amssymb,showpacs]{revtex4-1}
\usepackage{ulem}
\usepackage{amssymb}
\usepackage{graphicx}
\usepackage{bm}
\usepackage{subfigure}
\usepackage{epsfig}
\usepackage[ansinew]{inputenc}
\usepackage{color}
\usepackage{float}

\def\gtappr{{{\lower4pt\hbox{$>$} } \atop \widetilde{ \ \ \ }}}
\def\bk{{\bf k}}

\def\bR{{\bf R}}
\newcommand{\dg}{^{\dagger }}
\newcommand{\ybal}{$\beta$-YbAlB$_4\,$}

\newcommand{\yrs}{YbRh$_{2}$Si$_{2}\,$}

\graphicspath{ {figs/} }
\begin{document}

\title{Theory of the Electron Spin Resonance in the Heavy Fermion Metal \ybal}
 
\author{Aline Ramires}
\author{Piers Coleman}
\affiliation{Department of Physics and Astronomy, Rutgers University, Piscataway, New Jersey, 08854, USA}

\begin{abstract}
The
heavy fermion metal \ybal exhibits a bulk room temperature  conduction electron ESR signal which evolves into an Ising-anisotropic $f$-electron signal exhibiting hyperfine features at low temperatures. We develop a theory for this phenomenon based on the development of resonant scattering off a periodic array of Kondo centers. We show that the hyperfine structure arises from the scattering off the Yb atoms with nonzero nuclear spin, while the constancy of the ESR intensity is a consequence of the presence of crystal electric field excitations of the order of the hybridization strength.
\end{abstract}

\date{\today}
\maketitle

Heavy fermion systems have been probed by a variety of experimental
techniques and have provided great insights into the understanding of
strong correlated systems. These systems are formed by a lattice of
localized moments immersed in a conduction sea \cite{HF1,HF2}. An
important class of heavy fermion metals exhibits the phenomenon of
quantum criticality \cite{Si1,Si2}, and the recent discovery of an
intrinsically quantum critical heavy fermion metal, \ybal {\cite{Nak, Mat,Ram,Far}},
with an unusual electron spin resonance (ESR) signal \cite{Hol} has
attracted great interest.

Traditionally, ESR is used as a probe of
isolated magnetic ions in dilute rare-earth systems \cite{Bar}.  With the
discovery of sharp bulk ESR absorption lines in certain heavy fermion
materials, this experimental probe has emerged as a fascinating new
tool to probe the low energy paramagnetic spin fluctuations in these materials.
Normally, rare-earth ions display an  ESR signal when they are weakly coupled to the surrounding conduction sea, acting as dilute ``probe atoms". A bulk $f$-electron ESR signal in heavy fermion metals is unexpected, for here, the lattice of local moments are strongly coupled to the conduction electron environment. Naively, one expects the ESR resonance to be washed out by the Kondo effect, yet surprisingly, sharp ESR lines have been seen to develop at low temperatures in a variety of heavy electron
materials \cite{Kre, Koc}.

The case of \ybal, where the ESR signal evolves  from a room temperature conduction electron signal into an Ising-anisotropic $f$-electron signal at low temperatures, is particularly striking. As the temperature is lowered, the $g$ factor changes from an isotropic $g\approx 2$ to an anisotropic $g$ factor characteristic of the magnetic Yb ions. Moreover, the signal develops hyperfine satellites characteristic of localized magnetic moments, yet the intensity of the signal remains constant, a signature of Pauli paramagnetism \cite{Hol}. These results challenge our current understanding and motivate the development of a theory of spin resonance in the Anderson lattice.

Here, we formulate a phenomenological theory for the ESR of an Anderson
lattice containing anisotropic magnetic moments.  Our theory
builds on earlier works \cite{AW, Schl,Han}, focusing on the
interplay between the lattice Kondo effect and the paramagnetic spin
fluctuations while considering the effects of spin-orbit coupling,
crystal electric field (CEF) and hyperfine coupling. We show that the key features of the observed ESR
signal in \ybal, including the shift in the $g$ factor and the
development of anisotropy, can be understood as a result of the
development of a coherent many-body hybridization between the
conduction electrons and the localized $f$ states. We are able to
account for the emergence of hyperfine structure as a consequence of
the static Weiss field created by the nuclei of the odd-spin isotopes
of Yb. Moreover, using a spectral weight analysis, we show that 
the constancy of the intensity can be understood as a
consequence of the intermediate value of the CEF excitations, comparable to the hybridization strength.

ESR measurements probe the low frequency transverse magnetization
fluctuations in the presence of a static magnetic field. The power
absorbed from a transverse ac electromagnetic field at fixed frequency
$\nu_{0}$ as a function of the static external field $H$, is given by
\begin{equation}\label{}
P(\nu_{0},H)  \propto \chi'' _{+-} (\nu_{0},H),
\end{equation}
where 
\begin{equation}
\chi_{+-} (\nu,H) = -i \int_{0}^{\infty } dt e^{i\nu t} \langle [M_{+} (t),M_{-} (0)]\rangle_H
\end{equation}
is the dynamical transverse magnetic susceptibility and $M_{\pm}=M_{x}\pm i M_{y}$ are the raising and lowering components of the magnetization density. 

In \ybal, the Yb ions are sandwiched between two heptagonal rings of boron atoms \cite{Nak}, occupying a magnetic 4$f^{13}$ state with total angular momentum $J$=7/2. Crystal fields with sevenfold symmetry conserve $J_z$, splitting the $J$=7/2 Yb multiplet into four Kramers doublets, each with definite $|m_J|$. Based on the maximal degree of overlap, the Curie constant and the anisotropy of the magnetic susceptibility of \ybal, the low lying Yb doublet appears to be \hbox{$|m_J=\pm5/2\rangle$}, with first excited state $\vert m_{J}=\pm 3/2\rangle $ \cite{And, Lon}.

We start with an infinite-$U$ Anderson lattice model, based on the overlap of the boron orbitals with the $|7/2,\alpha=\pm5/2\rangle$ $f$-electron ground state doublet and the
first excited CEF level, the pure $|7/2,\beta=\pm3/2\rangle$ state, given by $H=H_{c}+H_{f}+H_{fc}-\mathbf{M}.\mathbf{H}$, where
\begin{eqnarray}
H_c&=& \sum_{\bk,\sigma} \epsilon_{\bk}c\dg_{\bk\sigma}c_{\bk\sigma},
\qquad  H_{f}=  \sum_{j,\gamma} \epsilon_{f\gamma}f\dg _{\gamma} (j) f_{\gamma} (j),
\cr
H_{fc} &=&\sum_{j\bk\sigma\gamma} ( e^{-i\bk\bR_j}
V_{\bk\sigma\gamma} c\dg_{\bk\sigma}
X_{0\gamma} (j)
+ {\rm
H.c.})
\end{eqnarray}
describe the conduction and $f$ bands, and the hybridization between them;  $\mathbf{M}= \sum_{j} \mu_B\biggl ( g_{c} \mathbf{S}_{c} (j)+ g_{f}\mathbf{J}_{f} (j)\biggr)$ is the total magnetization, where $g_{c}=2$ and $g_{f}= 8/7$ are the conduction and  $f$-electron
Land\'e $g$ factors and $\mathbf{J}_{f}$ is the total angular momentum operator of the 
$f$ states. The operator $c_{\bk\sigma}^\dagger$ creates a conduction hole in the boron band with dispersion $\epsilon_{\bk}$. The composite $X_{0\gamma}=(b\dg f_{\gamma})\equiv  \vert 4f^{14}\rangle \langle 4f^{13},\gamma \vert $ is the Hubbard operator between the $\vert  4f^{13}, \gamma \rangle \equiv f\dg_\gamma\vert  0 \rangle $, 
``hole''  states of the Yb$^{3+}$ ion and the filled shell Yb$^{2+}$ state $\vert 4f^{14}\rangle \equiv b\dg\vert 0\rangle $,  written using a slave boson representation.
The azimuthal quantum number $\gamma\equiv m_{J}$ has values $\gamma\in [\pm 5/2, \pm 3/2]$ corresponding to the ground state doublet with energy $\epsilon_{f\pm 5/2}=\epsilon_{f}$  and the next CEF level, with energy $\epsilon_{f\pm 3/2}=\epsilon_{f}+{\Delta_X}$.

We employ a mean-field approximation  $X_{0\gamma} (j)\rightarrow r f_{\gamma (j)}$, where the mean-field amplitude of the slave boson,  $r=|\langle b_{j} \rangle|$ 
describes the emergence of the
Abrikosov-Suhl resonance at each site, resulting from Kondo screening.  In the  mean field theory,  $H \rightarrow H_{c}+\tilde{H}_{f}+\tilde{ H}_{fc}$, where 
\begin{eqnarray}
\tilde{H}_{f}  &=& \sum_{\bk\gamma}\tilde{\epsilon}_{f\gamma} f_{\bk\gamma}\dg f_{\bk\gamma } +\lambda(r^2 -1),\\
\tilde{H}_{fc}&=& \sum_{\bk\sigma\gamma}
\bigl[ c\dg_{\bk\sigma} \tilde{V}_{\bk\sigma\gamma} 
f_{\bk\gamma }+ {\rm
h.c.}\bigr ],
\end{eqnarray}
with $\tilde{V}_{\bk\sigma\gamma}=V_{\bk\sigma\gamma} r$ and $\tilde{\epsilon}_{f\gamma}=\epsilon_{f\gamma}+\lambda$ the renormalized quasiparticle hybridization and $f$-level energy, and $\lambda$ the Lagrange multiplier that enforces the average constraint $\langle n_f \rangle + \langle b\dg b\rangle =1$.
The temperature dependence of the many body amplitude $r (T)$ and $\lambda(T)$ determines the evolution of the ESR signal. 

In the ground state, the ratio $\tilde{V}^{2}/W\sim T_{K}$ determines
the Kondo temperature $T_{K}$, where $\tilde{V}$ is the characteristic
size of the hybridization and $W$ is the conduction electron
bandwidth. The degree of magnetic anisotropy in the Kondo lattice is
set by the size of the crystal field splitting ${\Delta_X} $. In a Kondo impurity, one can project out the crystal field excited states, provided ${\Delta_X}/T_{K} \gtappr 1$, and crystal symmetry prevents any admixture of the projected states with the Abrikosov-Suhl resonance. However, in a Kondo lattice the nonconservation of crystal symmetry becomes important once $\Delta_{X} \gtappr \tilde{V}\sim \sqrt{T_{K}W}$, a situation that can occur even though $\Delta_{X}\gg T_{K}$. In this situation, the hybridization will admix the mobile $f$ quasiparticles with the higher crystal field states. We shall show that this produces significant modification to the magnetization operator of the quasiparticles. Thus, there are three regimes of interest:
\begin{enumerate}
\item Ising limit: $\Delta_{X} /\tilde{V}\gg 1$, ${\Delta_X}/T_{K}\gg 1$,
\item Intermediate anisotropy: $\Delta_{X}/\tilde{V}\gtappr 1$, ${\Delta_X}/T_K\gg 1$,
\item Weak anisotropy: $\Delta_{X}/\tilde{V}< 1$.
\end{enumerate}

Although \ybal almost certainly lies in the second category, 
the Ising limit captures most of the physics.
In this limit, the 
$\pm 3/2$ states are projected out, 
leading to a two-band
model in which the matrix elements of the 
transverse $f$ magnetization $J_{f}^{\pm} $ are
absent. 
The ESR signal, then, is determined by the spin
dynamics of the conduction electrons in the presence of the
lattice Kondo effect, given by $P (\nu,H)\propto
\chi''_{c+-} (\nu,H)$.  As a first step we
examine this limit, using a simplified model in which 
the hybridization is spin diagonal and its complex momentum dependence is ignored, 
replacing $\tilde{V}_{\bk\sigma \gamma}\rightarrow
\tilde{V}\underline{1}$. 
In mean-field theory, 
\begin{equation}\label{}
\chi_{c+-} (i\nu_n) = -  \mu_{B}^{2}T\sum_{m}
G_{c\downarrow  } (\bk,i\tilde{\omega}_m+i\nu_{n})
G_{c\uparrow } (\bk,i\tilde{\omega}_m)
\end{equation}
where  $G_{c\sigma } (z) =
\left[z-\epsilon_{\bk\sigma }-\Sigma_{c\sigma } (z)
\right]^{-1}$ is the conduction electron propagator and 
$\Sigma_{c\sigma } (z)= {V}^{2}r^{2}/
(z- \tilde{\epsilon}_{f\sigma })$ is the 
self-energy
generated by resonant scattering off $f$ states. Here vertex corrections have been neglected and the spin relaxation has been included as a white noise Weiss field acting on conduction and $f$ electrons, shifting the Matsubara frequency by the spin-relaxation
rate,  $\tilde{\omega}_{m}= \omega_{m}+ i \frac{\Gamma}{2}{\rm sgn} (\omega_{m})$. 
Carrying out the momentum sum as an energy integral, and expanding the self-energy to linear order in frequency, at low temperatures we obtain
\begin{equation}\label{Chi}
\chi_{c+-}(\nu-i\delta,H) =   \mu_{B}^{2}Z_{c}N_c(0)
\left(\frac{g^{*}\mu_{B }H+ i
\Gamma}{
g^{*}{\mu}_B H + i \Gamma -\nu
}
\right).
\end{equation}
Here, $N_c(0)$ is the density of states of the conduction electrons, $Z_{c}= (1- \partial \Sigma_c /\partial \omega)^{-1}= (1+V^{2}r^{2}/\tilde{\epsilon}_{f}^{2})^{-1}$ is the
conduction electron quasiparticle weight and 
\begin{equation}\label{}
g^{*}= g_{c}Z_{c}+  g_{f}^*(1-Z_{c})
\end{equation}
is the effective $g$ factor of the heavy quasiparticles at the Fermi surface (FS), where $g_f^*=g_f(2m_J) =5.7$.  At high temperatures, $g^{*}\approx 2$ reflects the conduction character of the FS, but as the temperature is lowered the $g$ factor rises towards $g_f^*$ as the FS acquires $f$ character. 
The evolution of $g^{*} (T)$, computed using the temperature-dependent mean-field parameters (Fig. 1), is qualitatively similar to that observed in \ybal,  but the asymptotic
value at low temperatures  is twice as large as that seen experimentally. Details of the computation can be seen in the Supplemental Material. In the Ising limit, the $f$ band responds uniquely to $z$-axis fields, so that when a field is applied at an angle $\theta $ from the plane perpendicular to the $z$ axis, we may decompose the $g$ factor in components parallel and perpendicular to the $c$ axis:
\begin{eqnarray}
g^*(\theta)&=&\sqrt{(g^{*}_{\parallel} \sin\theta)^2+ (g^{*}_{\perp} \cos\theta)^2}.
\end{eqnarray}
At high temperatures, $g^* (\theta ) = g_c$ is isotropic, but at low temperatures, $g^*
(\theta)\sim 5g_f\sin\theta$ exhibits Ising anisotropy (Fig. 1 inset). 

\begin{figure}[h]
\begin{center}
\includegraphics[width=0.85\linewidth, keepaspectratio]{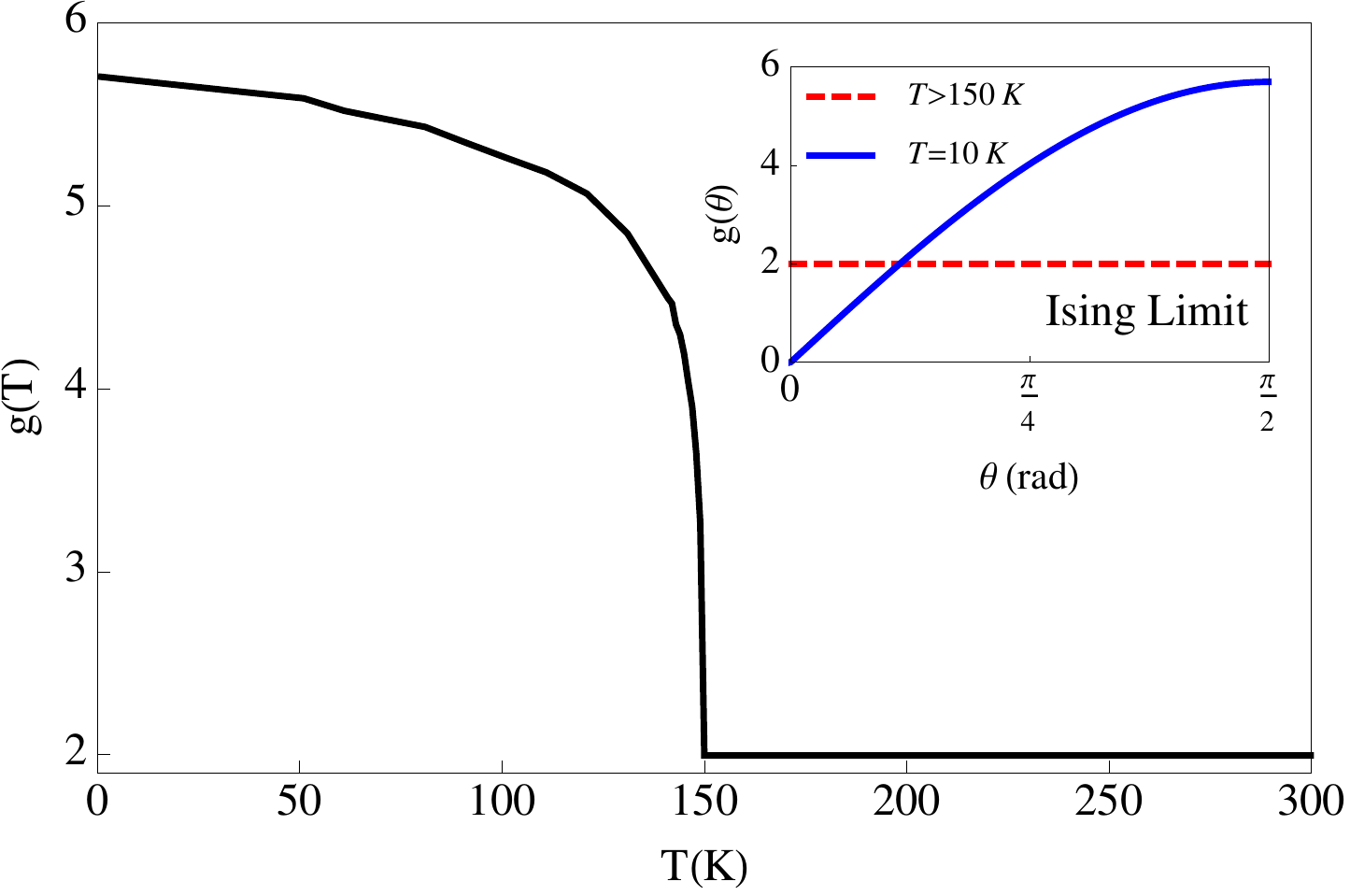}\label{FigG}
\caption{Temperature dependence of the thermally averaged  $g$ factor. The parameters used were $\epsilon_f=-0.15eV$ and $V=0.26eV$. See details of the mean field solution in the Supplemental Material. The inset shows the anisotropy of the $g$ factor in the Ising limit.}
\end{center}
\end{figure}

Next, we consider the effect of hyperfine coupling on the Kondo
lattice ESR signal. A small isotopic percentage ($n_{i}\approx $14\%)
of the Yb atoms in \ybal carry nuclear spins, which give rise to a
hyperfine coupling between the $f$ states and the nuclei \cite{Hol}.
The $f$ electrons at these sites experience a Weiss field of magnitude
$A$ that shifts the central energy $\tilde{\epsilon}_{f}$ of the Abrikosov-Suhl resonance. 
When we impurity average over the positions of the isotopic
impurities, this modifies the conduction electron self-energy
$\Sigma_{c\gamma} (z)\rightarrow \Sigma_{c\gamma} (z)+\delta
\Sigma_{c\gamma} (z)$, where 
\vspace{-0.25cm}
\begin{figure}[H]
\begin{center}
\begin{eqnarray}\label{l}
\delta \Sigma_{c\gamma }(z)&=&\raisebox{-3mm}[0cm][0cm]{
\includegraphics[width=0.75\linewidth,
keepaspectratio]{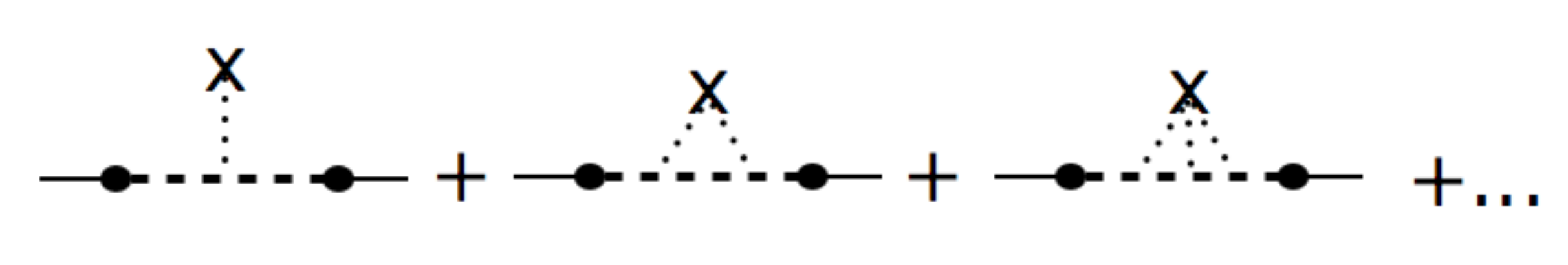}}\label{HFDiag}\cr\nonumber\\[-2pt]
&=&\frac{n_{i}\tilde{V}^{2}}{2}
\sum_{\sigma =\pm 1}
\left(
\frac{1}{z-\tilde{\epsilon}_{f\gamma }+A\sigma }
- 
\frac{1}{z- \tilde{\epsilon}_{f\gamma }} \right)
\end{eqnarray}
\end{center}
\end{figure}
\vspace{-0.6cm}
\noindent with the crosses representing the hyperfine field $A\sigma $ ($\sigma
=\pm 1$). The resulting electron self energy
\begin{eqnarray}
\hspace{-0.5cm}\Sigma_{c\gamma} (z )=\frac{(1-n_i)\tilde{V}^2}{z-\tilde{\epsilon}_{f\gamma }}+\frac{\frac{n_i}{2}\tilde{V}^2}{z-\tilde{\epsilon}_{f\gamma }+A}+\frac{\frac{n_i}{2}\tilde{V}^2}{z-\tilde{\epsilon}_{f\gamma}-A}
\end{eqnarray}
contains two extra resonances, shifted by the hyperfine coupling
constant $A$, which lead to two corresponding side peaks in the ESR
lines at low temperatures, as shown in Fig. 2.  Thus, we are able to
interpret the appearance of hyperfine peaks in the ESR signal of \ybal
as a consequence of the hyperfine splitting of the resonant scattering
in this Kondo lattice.

\begin{figure}[h]
\begin{center}
\includegraphics[width=0.85\linewidth, keepaspectratio]{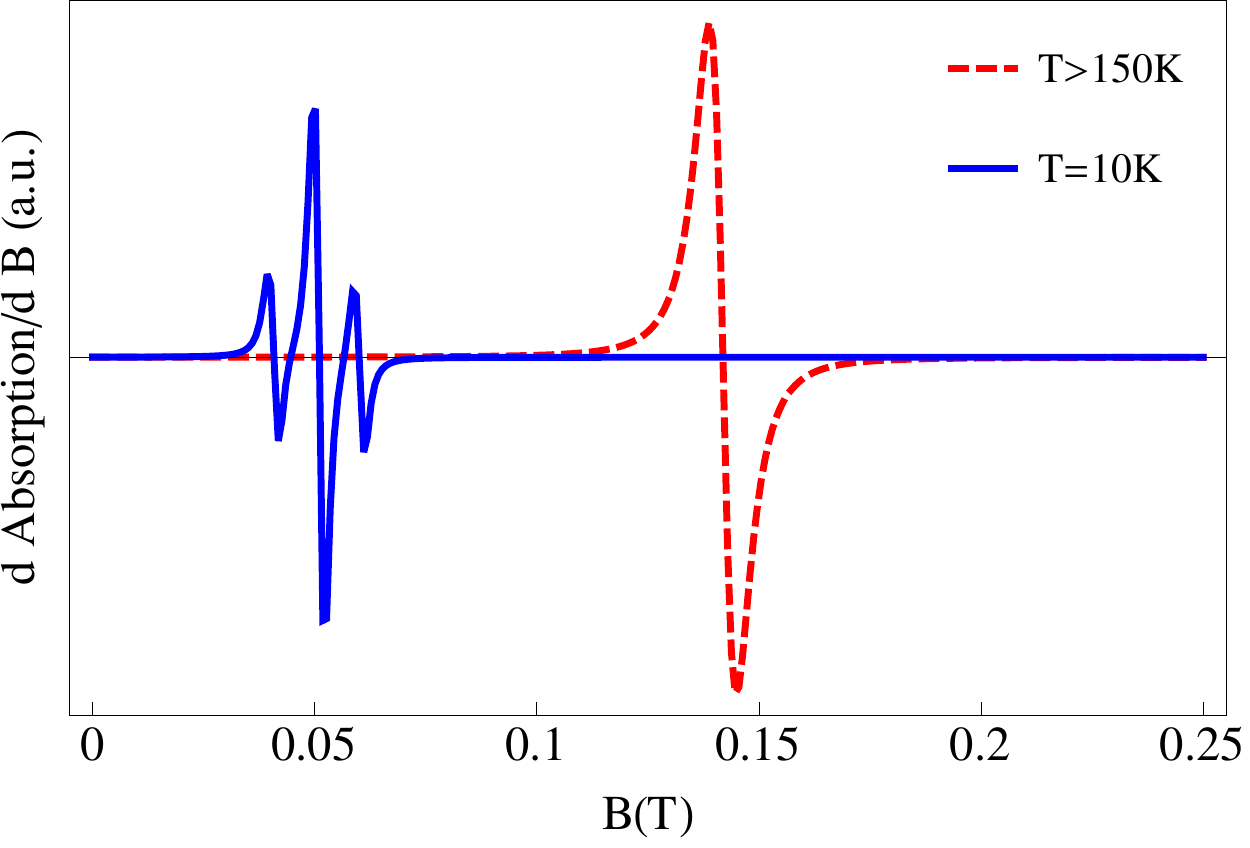}\label{HFESR}
\caption{ESR signal computed using the mean-field theory, 
including the effect of hyperfine coupling in the Abrikosov-Suhl resonance. Here $A=7.5 \times10^{-6}eV$ and $\Gamma=7.2\times10^{-7}eV$. The low temperature curve was rescaled by a factor of 10. Note the development of satellite peaks at low temperatures.}
\end{center}
\end{figure}

Now we turn to a discussion of the ESR
signal intensity  in \ybal.  
Here, we employ a sum rule relating the quasiparticle, 
or Pauli component of the magnetization to the ESR intensity. 
The ESR intensity is the field integral of the absorbed power, $I_{ESR}\propto \int_0^{H_{max}} 
\chi_{+-} '' (\nu_0,H) dH$, where $H_{max}$ is the maximum field applied
and $\nu_0$ is the fixed ESR frequency. We can write this in the form 
\begin{eqnarray}\label{sum1}
I_{ESR}\propto  H_0 
\int_0^{H_{max}} 
\frac{\chi_{+-} '' (\nu_0,H)}{\nu_0} g^*\mu_B 
dH,
\end{eqnarray}
where  
 $H_{0} = \nu_0/(2g^* \mu_{B})$  is the resonance field. 
Now since the integrand is an even  function of $\nu_0-2g^*\mu_B H$, it follows that $\chi_{+-} '' (\nu_0,H) = \chi ''_{+-}(\nu,H_0)$, where $\nu= 2g^{*}\mu_{B} H$. Writing 
$d\nu=2g^{*}\mu_{B}dH$, then
\begin{eqnarray}\label{intense}
I_{ESR}\propto  \frac{H_0}{2} \int_0^{\nu_{max}} 
\frac{\chi_{+-} '' (\nu,H_0)}{\nu} d\nu,
\end{eqnarray}
where $\nu_{max}=2g^*\mu_{B}H_{max}$ and we have used the narrowness of
the peak to replace $\nu_0\rightarrow \nu $ in the
denominator. There is also a sum rule for the total
transverse static susceptibility, given by the Kramers-Kr\"onig relation:
\begin{eqnarray}\label{static}
\chi_{+-}'(0,H_0)=\frac{1}{2\pi}
\int_{-\infty}^{\infty} \frac{\chi''_{+-} (\nu,H_0)}{\nu}d\nu.
\end{eqnarray}

In anisotropic $f$-electron systems like \ybal, the
transverse susceptibility is dominated by Van Vleck paramagnetism, and
is temperature independent. In this situation, (\ref{static})
plays the role of a magnetic $f$-sum rule. In fact, 
the static susceptibility $\chi ' (\nu=0,H_0)= \chi_{\rm
Pauli}+\chi_{VV}$ is a sum of Pauli and Van Vleck (VV)
susceptibilities, where the Pauli contribution derives from
low-frequency spin-flip processes, lying within the
frequency range detected by ESR, whereas the Van Vleck contributions
derive from much larger crystal-field frequencies.
In this way, we see that the ESR intensity measures the
Pauli component of the transverse magnetization,
\begin{eqnarray}
I_{ESR} (T)\propto 2 \pi H_0 \chi_{\rm Pauli} (T).
\end{eqnarray}

Experimentally, both the transverse static susceptibility
($\chi_{Total} (T) =\chi_{0}$, \cite{Mac}) and the ESR intensity
($I_{ESR}(T)=I_{0}$, \cite{Hol}) are
temperature independent. While the large constant value of the total susceptibility reflects its 
Van Vleck character, telling us that the total spectral weight  in
Eq. (\ref{static}) is conserved,
the temperature independence of the ESR
intensity means that the Pauli contribution to the spectral weight
is {\sl also} conserved. In the Ising limit, as the hybridization turns on, there is a large reduction in the conduction electron character of the FS, giving rise to a much reduced transverse magnetization and ESR intensity. Thus to account for these features we need to reinstate the finite CEF.

In the presence of a CEF level, the decomposition of the
quasiparticles into conduction and $f$ electrons  contains an additional
amplitude to be in the excited crystal field state $ |f_{3/2}\beta\rangle$,
\begin{equation}
|n \bk\sigma\rangle= a_{n\sigma} | c \bk\sigma\rangle+b_{n\alpha}|f_{5/2} \alpha\rangle + c_{n\beta} |f_{3/2}\beta\rangle.
\end{equation}
The low temperature Pauli part of the transverse susceptibility is written as $\chi_{Pauli}= N^{*} (0) |\langle 1 \bk \uparrow|M_+| 1 \bk \downarrow\rangle|^2 $, where $N^{*} (0)\sim 1/T_{K}$ is the low temperature quasiparticle density of states; thus, the ratio between the zero and room temperature intensities is given by 
\begin{equation}
\frac{I_{ESR} (0)}{I_{ESR} (T>T_K)} \propto \frac{N^*(0)}{N_c(0)} \frac{|\langle 1 \bk \uparrow|M_+| 1 \bk
\downarrow\rangle|^2 }{\mu_B^2},
\end{equation}
where $N_{c}(0)\sim 1/W$ is the conduction electron density of states and the matrix element at high temperatures is equal to $\mu_B^2$. The matrix element of the lower band ($n$=1) is $ |\langle 1 \bk\uparrow|M_+| 1 \bk\downarrow\rangle|^2 
=\mu_B^2|a_{1\uparrow}a_{1\downarrow}+g_f\sqrt{3}(b_{1\uparrow}c_{1\downarrow}+c_{1\uparrow}b_{1\downarrow})|^2$. Transitions between the $5/2$ and $3/2$ states happens via an
intermediate conduction state:
\begin{figure}[H]
\begin{center}
\includegraphics[width=0.6\linewidth, keepaspectratio]{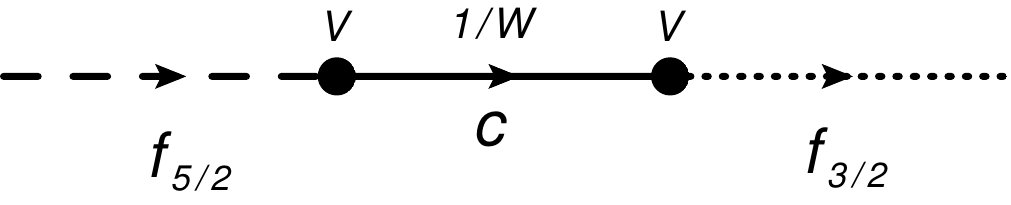}
\end{center}
\end{figure}
\vspace{-0.5cm}
\noindent
giving rise to a transition matrix element between the crystal field
states of magnitude $\tilde{V}^{2}/W \sim T_{K}$. The ground-state
quasiparticle amplitudes are, thus, of order
$(\sqrt{T_K/W},1,{T_K/{\Delta_X}})$, respectively.  In
the pure Ising limit ($\Delta_X \rightarrow \infty$) we have
$I_{ESR}(0)/I_{ESR} (T>T_K)\sim T_{K}/W\ll 1$ but at intermediate
anisotropy ($\Delta_{X}/\tilde{V}\gtappr 1$) new contributions to the
transverse magnetization appear and it acquires a value of order
unity, $I_{ESR}(0)/I_{ESR} (T>T_K)\sim
{W T_{K}}/{\Delta_{X}^{2}}
= (\tilde{V}/\Delta_{X})^{2}
\sim 1$.

The preservation of ESR intensity at low temperatures can also
be understood in terms of magnetic sum rules (Fig. 3). From Eq.( \ref{intense}), we see that the
ESR signal is a kind of ``magnetic Drude peak'' in
the dynamical spin susceptibility, slightly shifted from zero
frequency by the applied magnetic field. In a simple hybridization
model with Ising spins, there is a transfer of magnetic Drude weight to
high energies, a magnetic analog of the spectral weight transfer
which develops in the
optical conductivity \cite{Aep}. However, when a crystal field is
introduced, the transfer of spectral weight to high
energies is compensated by the downward transfer of spectral weight
from the crystal field levels due to admixture of $\pm 3/2$ states 
into the heavy bands. This
preserves a fraction of order $O (\tilde{V}/{\Delta_X} )^2$ of the low frequency
spectral weight.

\begin{figure}[h]
\includegraphics[width=0.9\linewidth, keepaspectratio]{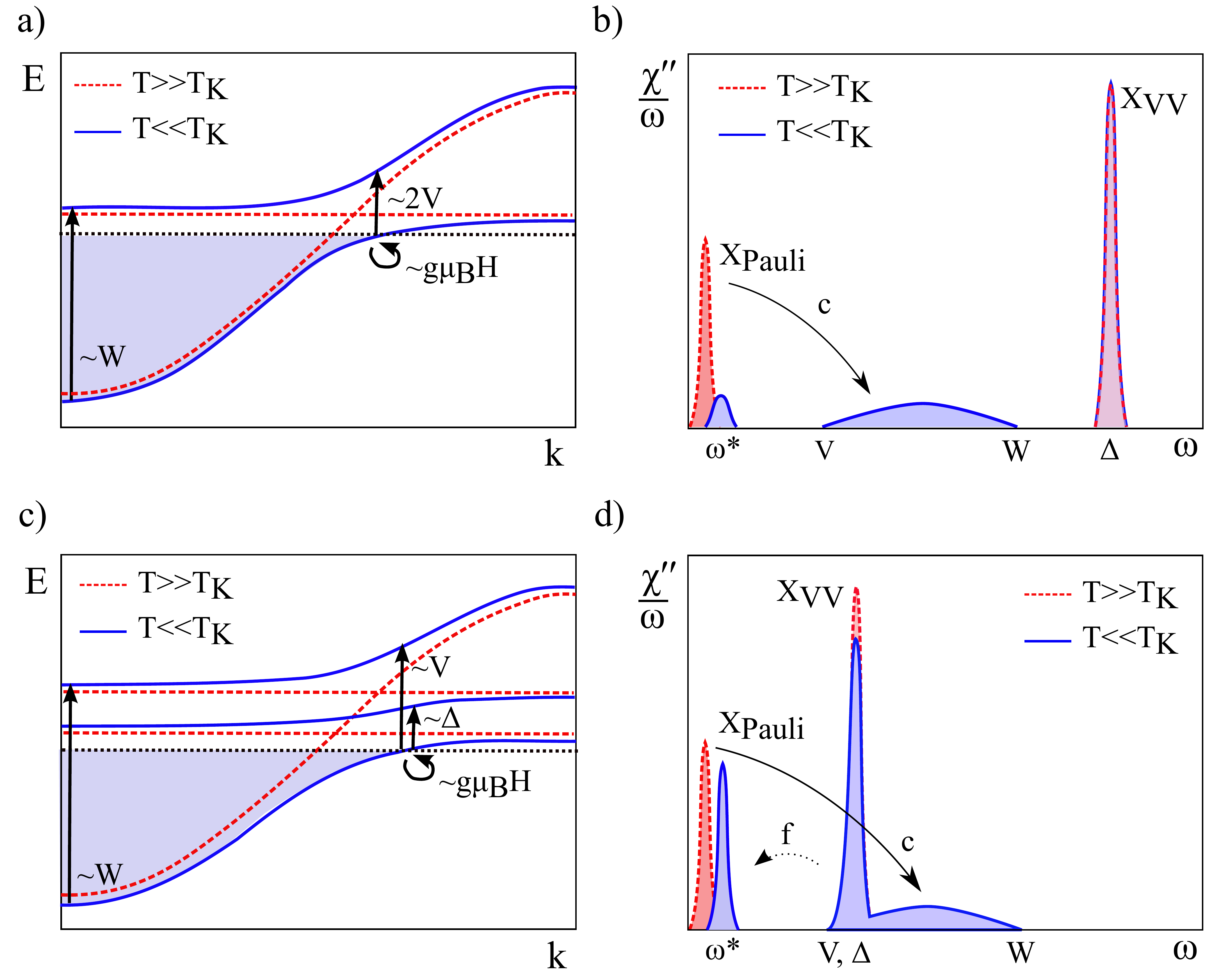}
\caption{Schematic plots of the bands a) Ising limit and c) Intermediate anisotropy. The arrows indicate the order of magnitude of the possible excitations. Imaginary part of the transverse spin susceptibility b) Ising limit and d) Intermediate anisotropy The arrows indicate the flow of spectral weight as the temperature is lowered.}
\end{figure}

Although we have not calculated it in detail, we note that the intermediate anisotropy limit allows us to understand the reduction of the ESR anisotropy. In particular, the momentum-space anisotropy of the
hybridization matrices $V_{\bk \sigma \gamma}$ will 
introduce a $k$-dependent rotation of the field quantization
axes. Quite generally, 
this effect will broaden the ESR line,
reducing  both the average value of the $g$ factor and the degree of
anisotropy of the signal.  

Our theory suggests various experiments to 
shed further light on our understanding of the spin paramagnetism of heavy
fermion systems. In particular, since \ybal is a Pauli limited
superconductor, we expect its upper critical field $H_{c2}$ to be
inversely proportional to the effective $g$ factor, so measuring the
angular dependence of $H_{c2}$ would allow us to independently confirm
the size and anisotropy of the $g$ factor.  It
would also be interesting to examine whether similar Ising anisotropic
systems, such as CeAl$_3$ or URu$_2$Si$_2$ and the quasicrystal YbAlAu \cite{Deg} exhibit ESR signals. Our
emergent hybridization model also raises many interesting
questions. For example, what is the underlying origin of the sharp
$f$-electron ESR line, which we have modeled phenomenologically? 
Moreover, is there a connection between the ESR resonance and
quantum criticality in both \ybal{\cite{Hol, Nak, Mat, Ram,Far}} and \yrs\cite{Sic1,Cus}? Tantalizingly, $\alpha$-YbAlB$_{4}$, a system with a structure locally similar to the $\beta$ phase does not
exhibit a $g$ shift, yet iron doping appears to drive it
into quantum criticality where a $g$ shift develops in the ESR \cite{Pas},
suggesting these two effects are closely
related. 
Clearly, these are issues for further investigation.


The authors would like to thank E. Abrahams, P. G. Pagliuso,
C. Rettori, R. R. Urbano, F. Garcia and L. M. Holanda
for discussions related to the ESR phenomena. This research was
supported by National Science Foundation Grants No. DMR-0907179 and No. DMR-1309929.

\section{Supplemental Material}

As mentioned in the main text, we employ a mean-field approximation  $X_{0\gamma} (j)\rightarrow r f_{\gamma } (j)$, where the mean-field amplitude of the slave boson,  $r=|\langle b_{j} \rangle|$  describes the emergence of the Abrikosov-Suhl resonance at each site, resulting from Kondo screening.  In the  mean field theory,  $H \rightarrow H_{c}+\tilde{H}_{f}+\tilde{ H}_{fc}$, where \begin{eqnarray}
H_c&=& \sum_{\bk,\sigma} \epsilon_{\bk}c\dg_{\bk\sigma}c_{\bk\sigma},\\
\tilde{H}_{f}  &=& \sum_{\bk\gamma}\tilde{\epsilon}_{f\gamma} f_{\bk\gamma}\dg f_{\bk\gamma } +\lambda(r^2 -1),\\
\tilde{H}_{fc}&=& \sum_{\bk\sigma\gamma}
\bigl[ c\dg_{\bk\sigma} \tilde{V}_{\bk\sigma\gamma} 
f_{\bk\gamma }+ {\rm
h.c.}\bigr ],
\end{eqnarray}
where $\tilde{V}_{\bk\sigma\gamma}=V_{\bk\sigma\gamma} r$ and $\tilde{\epsilon}_{f\gamma}=\epsilon_{f\gamma}+\lambda$ are the renormalized quasiparticle hybridization and f-level energy,  respectively. The Lagrange multiplier $\lambda$ enforces the average constraint $\langle n_f \rangle + \langle b\dg b\rangle =1$. 

Diagonalizing the mean field Hamiltonian within the assumption that the hybridization is momentum independent and spin-diagonal, one finds the energy bands:
\begin{equation}
E_{\bk\sigma}^\pm=\frac{\epsilon_{\bk\sigma}+\tilde{\epsilon}_{f\gamma}}{2}\pm\sqrt{\left(\frac{\epsilon_{\bk\sigma}-\tilde{\epsilon}_{f\gamma}}{2}\right)^2+\tilde{V}^2}.
\end{equation}

The temperature dependence of the ESR signal is determined by the temperature evolution of $r (T)$ and $\lambda(T)$. These are computed by the extremization the free energy, that can be written as:\begin{equation}
F=-\beta^{-1} \sum_{\bk ,\sigma,n} ln(1+e^{-\beta E_{\bk\sigma}^n})+\lambda(r^2-1),
\end{equation}
where $n=\pm$ is the band index and $\beta^{-1}=k_B T$.

The extremization of the free energy with respect to the mean field parameters $r$ and $\lambda$ are determined by:
\begin{eqnarray}
\frac{\partial F}{\partial r}=0, \hspace{1cm}
\frac{\partial F}{\partial \lambda}=0,
\end{eqnarray}
what gives two coupled equations:
\begin{eqnarray}
\sum_{\bk,\sigma,n} f(E_{\bk\sigma}^n)\frac{\partial E_{\bk \sigma}^n}{\partial r}+ 2\lambda r=0,\\
\sum_{\bk,\sigma,n} f(E_{\bk\sigma}^n)\frac{\partial E_{\bk \sigma}^n}{\partial \lambda}+ r^2-1=0
\end{eqnarray}
that are solved numerically.

For the numerical solution we use the equations above in a two dimensional square lattice with hopping parameter $t=1eV$, chemical potential $\mu=-0.2eV$. The location of the f-level is $\epsilon_f=-0.15eV$ and $V=0.26eV$. The temperature evolution of the mean field parameters is shown in the figure below:

\begin{figure}[H]
\begin{center}
\includegraphics[width=0.7\linewidth, keepaspectratio]{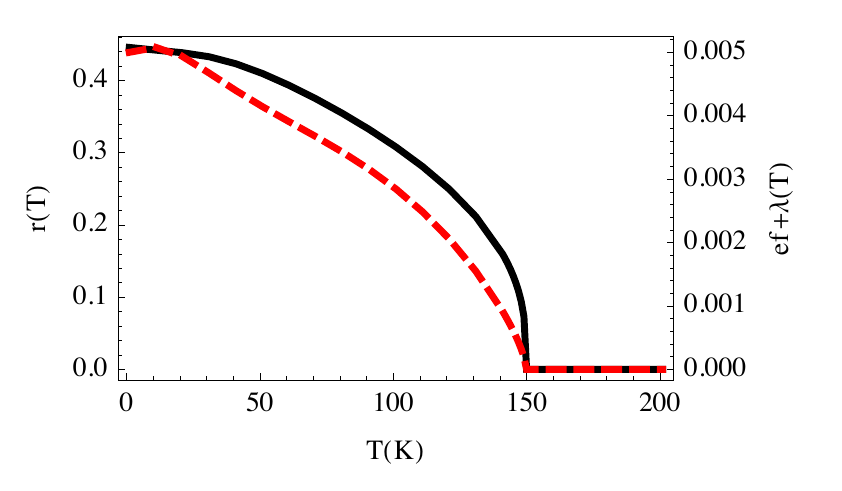}
\caption{Temperature dependence of the mean field parameters r (solid line) and $\lambda$ (dashed line) solved numerically.}
\end{center}
\end{figure}

In the main text we define $g^{*}= g_{c}Z_{c}+  g_{f}^*(1-Z_{c})$ as a simple closed form for the effective g-factor. This was a zero temperature calculation and gives only the value of the g-factor at the FS, what is a rough estimate of the real value of this parameter. 

We calculated the  temperature dependence of the g-factor from the
ratio between the photon energy $\nu_0$
and the Zeeman energy $2 \mu_B
H_{res}(T)$ at the resonance field, $g(T)=\nu_0/2 \mu_B H_{res}(T)$. 
$H_{res}$ was determined from the maximum 
of the imaginary part of the dynamical spin susceptibility 
\begin{equation}\label{}
\chi_{c+-} (\nu_{0}-i\delta,H) = -  \mu_{B}^{2}\hspace{-0.2cm}\sum_{\bk, n,m=\pm} \hspace{-0.2cm} Z_{\bk\downarrow}^{n}Z_{\bk\uparrow}^m\frac{ f(E_{\bk\downarrow}^n)-f(E_{\bk\uparrow}^m)}{-\nu_{0}+ E_{\bk\downarrow}^{n}-E_{\bk\uparrow}^m+i\Gamma},
\end{equation}
where,
\begin{equation}
Z_{\bk\sigma}^n=\frac{E_{\bk\sigma}^n-\epsilon_{f\sigma}}{E_{\bk\sigma}^n-E_{\bk\sigma}^{-n}},
\end{equation}
calculated at fixed frequency, as a function of field. 
In our actual calculations 
we used the experimental value for
the fixed ESR frequency $\nu_0=3.9\times 10^{-5}$eV (for the X-band frequency of $9.5$ GHz) and
$\Gamma=7.2\times10^{-7}eV$. The numerical solution is plotted in
Fig.1 in the main text.

\end{document}